\documentclass[twocolumn,showpacs,floatfix,secnumarabic,amssymb,
amsmath,aps,prl,nobibnotes,superscriptaddress,letter]{revtex4}

\usepackage{revsymb}
\usepackage{graphicx}
\usepackage{dcolumn}
\usepackage{bm}
\usepackage{epstopdf}
\usepackage{color} 

\newcommand{\Npart}{$\langle N_{part}\rangle$}
\newcommand{\Nparts}{$\langle N_{part}\rangle $ }
\newcommand{\sNN}[1]{$\sqrt{s_{NN}} = #1$ GeV}
\newcommand{\Lam}{$\Lambda$ }
\newcommand{\ALam}{$\bar{\Lambda}$ }
\newcommand{\Kaon}{$K^{0}_{S}$ }

\begin{document}

\title{Strangeness Enhancement in Cu+Cu and Au+Au Collisions at \sNN{200}}

\affiliation{Argonne National Laboratory, Argonne, Illinois 60439, USA}
\affiliation{University of Birmingham, Birmingham, United Kingdom}
\affiliation{Brookhaven National Laboratory, Upton, New York 11973, USA}
\affiliation{University of California, Berkeley, California 94720, USA}
\affiliation{University of California, Davis, California 95616, USA}
\affiliation{University of California, Los Angeles, California 90095, USA}
\affiliation{Universidade Estadual de Campinas, Sao Paulo, Brazil}
\affiliation{University of Illinois at Chicago, Chicago, Illinois 60607, USA}
\affiliation{Creighton University, Omaha, Nebraska 68178, USA}
\affiliation{Czech Technical University in Prague, FNSPE, Prague, 115 19, Czech Republic}
\affiliation{Nuclear Physics Institute AS CR, 250 68 \v{R}e\v{z}/Prague, Czech Republic}
\affiliation{University of Frankfurt, Frankfurt, Germany}
\affiliation{Institute of Physics, Bhubaneswar 751005, India}
\affiliation{Indian Institute of Technology, Mumbai, India}
\affiliation{Indiana University, Bloomington, Indiana 47408, USA}
\affiliation{Alikhanov Institute for Theoretical and Experimental Physics, Moscow, Russia}
\affiliation{University of Jammu, Jammu 180001, India}
\affiliation{Joint Institute for Nuclear Research, Dubna, 141 980, Russia}
\affiliation{Kent State University, Kent, Ohio 44242, USA}
\affiliation{University of Kentucky, Lexington, Kentucky, 40506-0055, USA}
\affiliation{Institute of Modern Physics, Lanzhou, China}
\affiliation{Lawrence Berkeley National Laboratory, Berkeley, California 94720, USA}
\affiliation{Massachusetts Institute of Technology, Cambridge, MA 02139-4307, USA}
\affiliation{Max-Planck-Institut f\"ur Physik, Munich, Germany}
\affiliation{Michigan State University, East Lansing, Michigan 48824, USA}
\affiliation{Moscow Engineering Physics Institute, Moscow Russia}
\affiliation{NIKHEF and Utrecht University, Amsterdam, The Netherlands}
\affiliation{Ohio State University, Columbus, Ohio 43210, USA}
\affiliation{Old Dominion University, Norfolk, VA, 23529, USA}
\affiliation{Panjab University, Chandigarh 160014, India}
\affiliation{Pennsylvania State University, University Park, Pennsylvania 16802, USA}
\affiliation{Institute of High Energy Physics, Protvino, Russia}
\affiliation{Purdue University, West Lafayette, Indiana 47907, USA}
\affiliation{Pusan National University, Pusan, Republic of Korea}
\affiliation{University of Rajasthan, Jaipur 302004, India}
\affiliation{Rice University, Houston, Texas 77251, USA}
\affiliation{Universidade de Sao Paulo, Sao Paulo, Brazil}
\affiliation{University of Science \& Technology of China, Hefei 230026, China}
\affiliation{Shandong University, Jinan, Shandong 250100, China}
\affiliation{Shanghai Institute of Applied Physics, Shanghai 201800, China}
\affiliation{SUBATECH, Nantes, France}
\affiliation{Texas A\&M University, College Station, Texas 77843, USA}
\affiliation{University of Texas, Austin, Texas 78712, USA}
\affiliation{University of Houston, Houston, TX, 77204, USA}
\affiliation{Tsinghua University, Beijing 100084, China}
\affiliation{United States Naval Academy, Annapolis, MD 21402, USA}
\affiliation{Valparaiso University, Valparaiso, Indiana 46383, USA}
\affiliation{Variable Energy Cyclotron Centre, Kolkata 700064, India}
\affiliation{Warsaw University of Technology, Warsaw, Poland}
\affiliation{University of Washington, Seattle, Washington 98195, USA}
\affiliation{Wayne State University, Detroit, Michigan 48201, USA}
\affiliation{Institute of Particle Physics, CCNU (HZNU), Wuhan 430079, China}
\affiliation{Yale University, New Haven, Connecticut 06520, USA}
\affiliation{University of Zagreb, Zagreb, HR-10002, Croatia}

\author{G.~Agakishiev}\affiliation{Joint Institute for Nuclear Research, Dubna, 141 980, Russia}
\author{M.~M.~Aggarwal}\affiliation{Panjab University, Chandigarh 160014, India}
\author{Z.~Ahammed}\affiliation{Variable Energy Cyclotron Centre, Kolkata 700064, India}
\author{A.~V.~Alakhverdyants}\affiliation{Joint Institute for Nuclear Research, Dubna, 141 980, Russia}
\author{I.~Alekseev~~}\affiliation{Alikhanov Institute for Theoretical and Experimental Physics, Moscow, Russia}
\author{J.~Alford}\affiliation{Kent State University, Kent, Ohio 44242, USA}
\author{B.~D.~Anderson}\affiliation{Kent State University, Kent, Ohio 44242, USA}
\author{C.~D.~Anson}\affiliation{Ohio State University, Columbus, Ohio 43210, USA}
\author{D.~Arkhipkin}\affiliation{Brookhaven National Laboratory, Upton, New York 11973, USA}
\author{G.~S.~Averichev}\affiliation{Joint Institute for Nuclear Research, Dubna, 141 980, Russia}
% ...

\collaboration{STAR Collaboration}
\noaffiliation

\date{\today}
\begin{abstract}

We report new STAR measurements of mid-rapidity yields for the $\Lambda$, $\bar{\Lambda}$, $K^{0}_{S}$, $\Xi^{-}$, $\bar{\Xi}^{+}$, $\Omega^{-}$, $\bar{\Omega}^{+}$ particles in Cu+Cu collisions at \sNN{200}, and mid-rapidity yields for the $\Lambda$, $\bar{\Lambda}$, $K^{0}_{S}$ particles in Au+Au at \sNN{200}. We show that at a given number of participating nucleons, the production of strange hadrons is higher in Cu+Cu collisions than in Au+Au collisions at the same center-of-mass energy. We find that aspects of the enhancement factors for all particles can be described by a parameterization based on the fraction of participants that undergo multiple collisions.

\end{abstract}

\maketitle

Relativistic heavy-ion collisions aim to create the QGP (Quark-Gluon Plasma), a unique state of matter where quarks and gluons can move freely over large volumes in comparison to the typical size of a hadron. Measurements of strangeness enhancement in heavy-ion collisions were originally conceived to be a key signature of QGP formation \cite{RafalMull}. It was argued that due to a drop in the strange quark's dynamical mass, strangeness in the QGP would equilibrate on small time scales relative to those in a hadronic gas \cite{StrangeHadron}. Assuming a thermally equilibrated QGP hadronizes into a maximum entropy state, a test for strange quark saturation in the early stages is provided by comparing final state hadron yields to thermal model predictions from the canonical formalism \cite{CanoncialSuppression}. These predictions have qualitatively reproduced various aspects of the data from Au+Au \sNN{200} collisions at RHIC (Relativistic Heavy Ion Collider); however, as with SPS (Super Proton Synchrotron) energies, a complete theoretical description has yet to be achieved \cite{HelensPRC}. We present mid-rapidity strange particle yields from Cu+Cu and Au+Au \sNN{200} collisions. Measurements at the AGS (Alternating Gradient Synchrotron) showed $K^{+}$ and $K^{-}$ yields to be higher in lighter systems compared to the respective values in heavy systems at a given number of participants \cite{AGS}. Measurements at the SPS showed higher $K/ \pi$ ratios for the light systems also at a given number of participants \cite{SPS}. Whether these trends continue up to RHIC energies, and what new information can be learned from strangeness enhancement as a QGP signature at RHIC, will be central issues in this Letter.

\begin{figure}[t]
\begin{center}
\includegraphics[width = 0.48\textwidth]{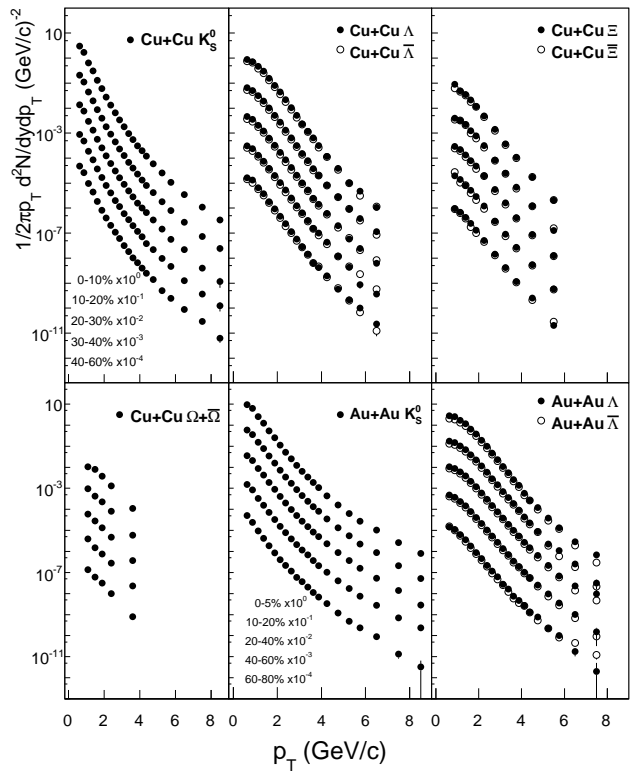}
\end{center}
\caption{$K^{0}_{S}$, $\Lambda$, $\bar{\Lambda}$, $\Xi$, $\bar{\Xi}$, and $\Omega+\bar{\Omega}$ invariant mass spectra from Cu+Cu and Au+Au \sNN{200} collisions, where $|y| < 0.5$. The $\Lambda$ and $\bar{\Lambda}$ yields have not been feed down subtracted from weak decays. The uncertainties on the spectra points are statistical and systematic combined.}
\label{Fig0}
\end{figure}

The new data presented are from approximately 20 million Au+Au \sNN{200} and 40 million Cu+Cu \sNN{200} collisions recorded at RHIC in 2004 and 2005, respectively. In order to extract the $\Lambda$, $\bar{\Lambda}$, $K^{0}_{S}$, $\Xi^{-}$, $\bar{\Xi}^{+}$, $\Omega^{-}$, $\bar{\Omega}^{+}$ yields as a function of transverse momentum, $p_T$, STAR's \cite{STAROver} Time Projection Chamber (TPC) \cite{STARTPC} is utilized to identify these particles via their dominant weak decay channels. The channels are $\Lambda \rightarrow p+\pi^{-}$, $\bar{\Lambda} \rightarrow \bar{p}+\pi^{+}$, $K^{0}_{S} \rightarrow \pi^{+}+\pi^{-}$, $\Xi^{-} \rightarrow \Lambda+\pi^{-}$, $\bar{\Xi^{+}} \rightarrow \bar{\Lambda}+\pi^{+}$, $\Omega \rightarrow \Lambda+K^{-}$, and $\bar{\Omega} \rightarrow \bar{\Lambda}+K^{+}$. These particles usually decay before the TPC's inner radius (50 cm), so the decay products enter the TPC.Daughter tracks are then reconstructed using STAR's tracking software. The raw particle yields are then calculated from the respective invariant mass distributions formed by the daughter track candidates. A combination of topological, energy loss, and kinematic restrictions are placed to ensure the combinatorial background is minimal, while preserving the statistical significance of the signal. We fit the regions adjacent to the respective peaks with a 2nd order polynomial, to determine the background beneath the respective peaks. This is then subtracted to obtain the signal. The signal to background ratio varies from 1 to 50, and depends on particle type, $p_T$, and the average charged particle multiplicity. To calculate the reconstruction efficiency, Monte Carlo particles are generated, embedded in the real events and propagated through a detector simulation. \textcolor{black}{The $ $\Lam and $ $\ALam yields have contributions from weak decays of charged and neutral $\Xi$ and their anti-particles, which can be subtracted up to $p_{T} \sim 5$ GeV/{\it c}.  This contribution is $\sim15\%$ and independent of $p_{T}$.} Feed-down contributions from $\Omega$ hadrons are negligible. More detailed descriptions of the strange particle spectra extraction can be found elsewhere \cite{Mythesis, STARpp}. The systematic uncertainties are due to: 1) slight mismatches in the real and embedded particle distributions which leads to an uncertainty in the reconstruction efficiency ($2-11\%$), and 2) small variations in raw particle yields with respect to the magnetic field setting and day ($\sim 2 \%$). Some of these uncertainties are common for Cu+Cu and Au+Au spectra. Finally, for each colliding system, data are partitioned in centrality bins, based on the charged hadron multiplicity in the pseudorapidity range $|\eta| < 0.5$.

\begin{figure}[t]
\begin{center}
\includegraphics[width = 0.48\textwidth]{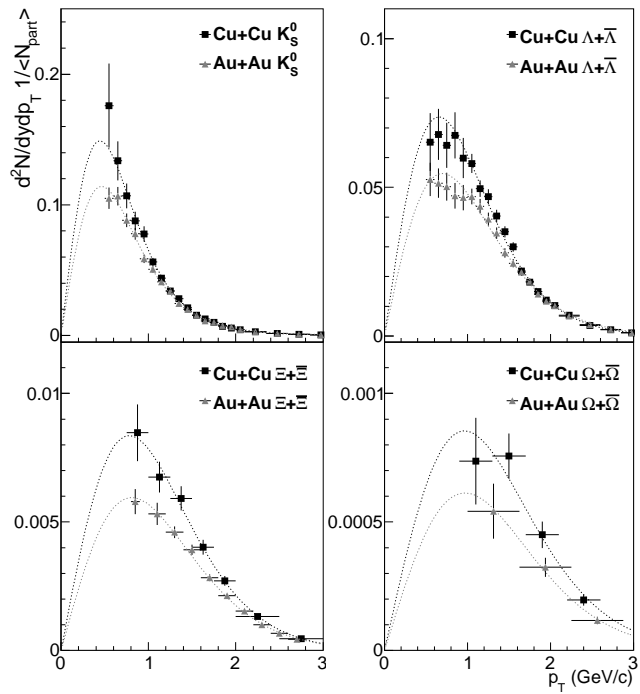}
\end{center}
\caption{$K^{0}_{S}$, $\Lambda+\bar{\Lambda}$, $\Xi+\bar{\Xi}$, and $\Omega+\bar{\Omega}$ spectra divided by \Npart$ $ for Cu+Cu $0-10\%$ ($\langle N_{part}\rangle\sim 99$) and Au+Au $20-40\%$  ($\langle N_{part}\rangle\sim 141$)  \sNN{200} collisions,  where $|y| < 0.5$. The Au+Au multi-strange data have been previously published \cite{AuAuLambda}. The $\Lambda$ and $\bar{\Lambda}$ yields have been feed down subtracted from weak decays. The uncertainties on the spectra points are statistical and systematic; for clarity the uncertainty on \Npart$ $ has not been included. The curves show the functions described in the text used to extract $dN/dy$.}
\label{Fig1}
\end{figure}
\begin{table*}
\begin{ruledtabular}
\begin{tabular}{lcccccc}
Cu+Cu & 0-10\% & 10-20\% & 20-30\% & 30-40\% & 40-60\% \\
\Npart & 99.0$\pm{1.5}$ & 74.6$\pm{1.2}$ & 53.7$\pm{1.0}$  & 37.8$\pm{0.7}$ &  21.5$\pm{0.5}$ \\
\hline $K^{0}_{S}$ & 13.9$\pm{1.0}$ & 9.81$\pm{0.68}$ & 6.49$\pm{0.44}$ & 4.22$\pm{0.32}$ & 2.24$\pm{0.23}$ \\
$\Lambda$ & 4.68$\pm{0.45}$ & 3.20$\pm{0.31}$ & 2.13$\pm{0.21}$ & 1.40$\pm{0.14}$ & 0.72$\pm{0.07}$ \\
$\bar{\Lambda}$ & 3.79$\pm{0.37}$ & 2.60$\pm{0.25}$ & 1.75$\pm{0.17}$ & 1.16$\pm{0.11}$ & 0.60$\pm{0.06}$ \\

$\Xi$ & 0.62$\pm{0.08}$ & 0.35$\pm{0.04}$ & 0.23$\pm{0.03}$ & 0.15$\pm{0.02}$ & 0.08$\pm{0.01}$  \\
$\bar{\Xi}$ & 0.52$\pm{0.08}$ & 0.32$\pm{0.046}$ & 0.20$\pm{0.03}$ & 0.16$\pm{0.03}$ & 0.07$\pm{0.01}$ \\
$\Omega+\bar{\Omega}$ & 0.141$\pm{0.017}$ & 0.106$\pm{0.012}$ & 0.068$\pm{0.008}$ & 0.045$\pm{0.007}$ & 0.015$\pm{0.003}$ \\ 
\hline 
Au+Au & 0-5\% & 10-20\% & 20-40\% & 40-60\% & 60-80\% \\
\Npart & 350$\pm{4}$ & 238$\pm{5}$& 147$\pm{4}$  & 67.5$\pm{2.7}$ & 23.0$\pm{1.2}$\\
\hline $K^{0}_{S}$ & 43.5$\pm{2.4}$ & 27.8$\pm{1.4}$ & 16.5$\pm{0.83}$ & 7.26$\pm{0.49}$ & 2.14$\pm{0.19}$ \\
$\Lambda$ & 14.8$\pm{1.5}$ & 9.16$\pm{0.89}$ & 5.70$\pm{0.55}$ & 2.38$\pm{0.23}$ & 0.71$\pm{0.07}$ \\
$\bar{\Lambda}$ & 11.7$\pm{0.9}$ & 7.27$\pm{0.55}$ & 4.53$\pm{0.34}$ & 1.82$\pm{0.14}$ & 0.55$\pm{0.04}$ \\

\end{tabular}
\caption{Mid-rapidity $dN/dy$ for strange hadrons in Cu+Cu and Au+Au \sNN{200} collisions. Combined statistical and systematic errors are shown}
\label{Tab:dNdy}
\end{ruledtabular}
\end{table*} 

Figure \ref{Fig0} shows the $p_T$ spectra for the singly-strange and multi-strange particles. A L\'evy function is used in this analysis to fit the spectra in order to extrapolate to the unmeasured region \cite{Levy}, so that the yield, $dN/dy$, can be extracted (see table \ref{Tab:dNdy}).
%
%
%
Uncertainties resulting from the extrapolation procedure, based on the above fit function, are included in the systematic uncertainties. Fits to the spectra for a selection of centralities are shown in fig. \ref{Fig1} on a linear scale. \textcolor{black}{The Au+Au \Kaon spectra were found to be consistent with published STAR $\langle K^{\pm} \rangle$ spectra \cite{ChargedKS}. We also found the Au+Au \Kaon spectra to be consistent with PHENIX and BRAHMS $\langle K^{\pm} \rangle$ spectra, apart from the very peripheral PHENIX data \cite{ChargedKB, ChargedKP, LongPaper}.} 

\begin{figure}[h]
\begin{center}
\includegraphics[width = 0.48\textwidth]{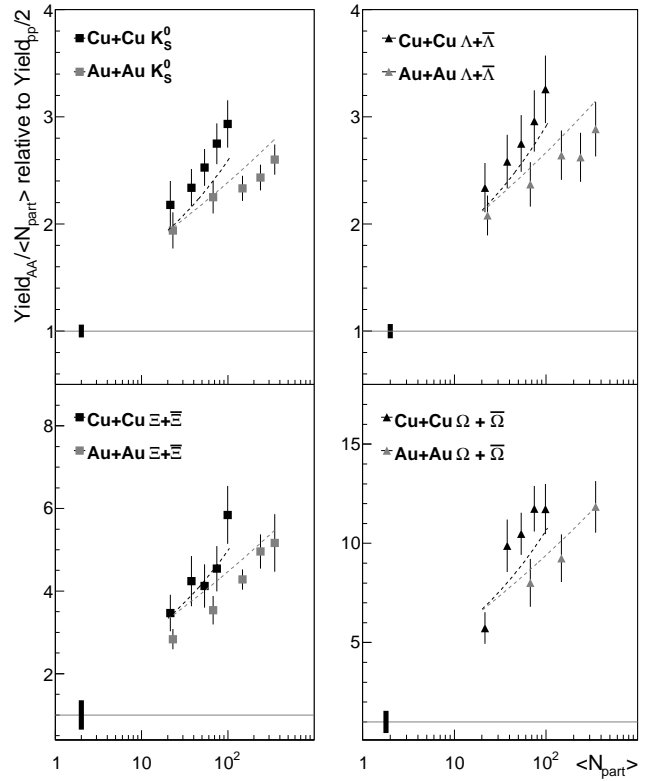}
\end{center}
\caption{The enhancement factor for (multi-) strange particles in Cu+Cu and Au+Au \sNN{200} collisions, where $|y| < 0.5$. The $\Lambda$ and $\bar{\Lambda}$ yields have been feed-down subtracted in all cases. The Au+Au multi-strange data have been previously published \cite{AuAuLambda}. The black bars show the normalization uncertainties, and the uncertainties for the heavy-ion points are the combined statistical and systematic errors. Curves described in the text, where $B_{K}= 2.0$, $B_{\Lambda}= 2.4$, $B_{\Xi}= 5.0$ and $B_{\Omega}= 12.1$.}
\label{Fig2}
\end{figure}

\begin{figure}[h]
\begin{center}
\includegraphics[width = 0.4\textwidth]{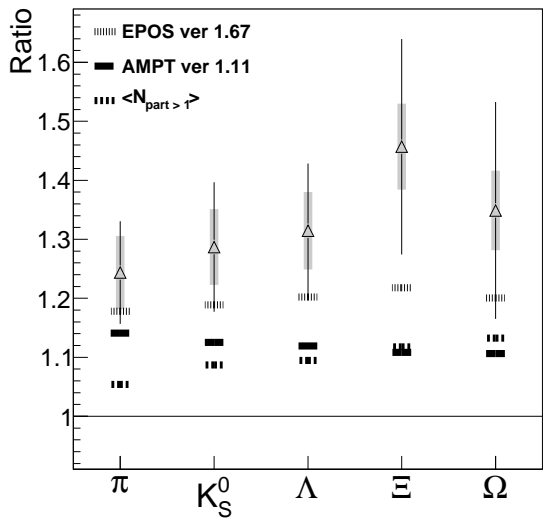}
\end{center}
\caption{Ratio of particle yields in central Cu+Cu and mid-central Au+Au collisions when $\langle N_{part} \rangle  =99$ in each case for $|y| < 0.5$. $\pi$ yields are from elsewhere \cite{AnetaPri}. Boxed uncertainties are from the Glauber calculations and are correlated for every particle. $\langle N_{part > 1} \rangle $ refers to the parameterization shown by eq. \ref{equ:fraction}, while the EPOS and AMPT models are described in the text. The default settings are used for each model. The vertical lines show the remaining independent statistical and systematic uncertainties.}
\label{Fig4}
\end{figure}

The enhancement factor, $E$, is defined as $dN/dy$ (yield) per mean number of nucleon participants ($\langle N_{part} \rangle$) in heavy-ion collisions, divided by the respective value in p+p collisions \cite{STARpp}. It characterizes the deviation in participant scaled yields relative to p+p.  Monte Carlo Glauber calculations are used to calculate $\langle N_{part} \rangle$ for each centrality bin in heavy-ion collisions \cite{LongPaper}. The top panels of fig. \ref{Fig2} show the enhancement factor for singly (anti-) strange particles in Cu+Cu and Au+Au collisions as a function of $\langle N_{part} \rangle$. In addition to the rising enhancements exhibited by all particles for both Cu+Cu and Au+Au collisions, at a given value of \Nparts above $\sim60$, the production of strange hadrons is higher in Cu+Cu collisions than in Au+Au collisions. Similar patterns are observed for the multi-strange particles in the bottom panels of fig. \ref{Fig2}.  \textcolor{black}{The Cu+Cu and Au+Au difference also applies to the non-strange sector, as shown in fig. \ref{Fig4}.} Finally, as shown in fig. \ref{Fig1}, the higher yields per $\langle N_{part} \rangle$ in Cu+Cu apply across the measured $p_{T}$ range, $p_{T} > 0.5$ GeV.
\textcolor{black}{It is assumed in the canonical framework that the observed strangeness enhancement actually results from a suppression of strangeness production in p+p collisions \cite{CanoncialSuppression}. This suppression arises from the need to conserve strangeness within a small, local volume, which limits strangeness production in p+p relative to A+A collisions. The correlation volume is a parameter in the canonical model which dictates the region to which strangeness conservation applies.} Assuming the system's correlation volume is proportional to \Npart, the canonical framework predicts yields per \Nparts which should rise with increasing \Nparts as phase space restrictions due to strangeness conservation are lifted.  At the grand canonical limit where $\langle N_{part} \rangle \sim 100$, yields per \Nparts are constant as a function of \Npart. The extracted chemical freeze-out temperature ($T_{ch}$) and baryo-chemical potential ($\mu_{b}$) values for Cu+Cu and Au+Au which are explicitly used for the framework's predictions, have been shown to be consistent and independent of system size \cite{AnetaPri}. Therefore, the higher yields in Cu+Cu and the rising Au+Au enhancements with $\langle N_{part} \rangle > 100$ in fig. \ref{Fig2} appear inconsistent with the canonical framework as the sole description of strangeness enhancement. There are other canonical predictions which assume the correlation volume may scale with $\langle N_{part} \rangle^{1/3}$ or $\langle N_{part} \rangle^{2/3}$ and these give slower rises of $E$ as a function of \Npart$ $ \cite{MLamontSQM}. Although these match the Au+Au data better, they also predict the enhancement should just depend on \Nparts which is again inconsistent with the Cu+Cu and Au+Au data. \textcolor{black}{If the canonical formalism is valid in describing strangeness enhancement, these failures may relate to the validity of the assumption that the correlation volume is proportional  to \Npart.}

The curves in fig. \ref{Fig2} correspond to the following parameterization:
\begin{equation}
\label{equ:fraction}
E_{i}(N_{part}) = B_{i}f(N_{part})+1
\end{equation}
%
which Becattini and Manninen (BM) propose as a core-corona description of strangeness production in heavy-ion collisions \cite{phiCoreCorona1}. The variable $f$ is the fraction of participants that undergo multiple collisions obtained from the Glauber model, and $B_{i}$ is a particle-wise normalisation factor. In this case, it is chosen to fit the Cu+Cu and Au+Au data simultaneously, therefore independent of collision species. Participants that undergo multiple collisions produce a core that expands and freezes out to produce hadrons. The resulting strange hadron yields follow thermal expectations for the reasons stated in the introduction of this Letter; namely that $s+\bar{s}$ equilibrate in the core's QGP stage, then the core hadronizes to produce strange hadrons in chemical equilibrium. $B_{i}$ depends linearly on the particle density in the core. Participants with just one collision act like nucleons in N+N collisions with respect to strangeness production. 

The parameterization describes the two main qualitative aspects of the data: the rising enhancements with \Npart$ $ in a given system over the full range of \Npart, and a higher enhancement factor for central Cu+Cu collisions compared to Au+Au collisions with the same \Npart$ $. The higher $E$ for Cu+Cu at a given \Npart$ $ simply results from $f(N_{part})$ being higher for the lighter system. This in turn is due to the differing geometries of the respective collision zones, i.e. Cu+Cu is more spherical at a given \Npart$ $. Although not implicit in the Glauber model, differing nuclear shadowing in Cu+Cu compared to Au+Au could also lead to larger multiple interactions in Cu+Cu at a given \Npart$ $\cite{Shadowing}. $f(N_{part})$ increases with centrality for a given system because the participant densities in the collision zone increase.  \textcolor{black}{Its important to note deviations from the curves are observed for the singly strange particles in central Au+Au and multi-strange particles in peripheral Au+Au multi-strange particles.  Since for a given particle, since we adjust $B_i$ in eq. \ref{equ:fraction} to best fit the Cu+Cu and Au+Au enhancements simultaneously, this sometimes leads to a poorer description of the Au+Au enhancements in relation to what is shown by BM where the Au+Au data alone is fit \cite{phiCoreCorona1}. As will be shown in fig. \ref{Fig4}, the relative differences in central Cu+Cu and mid-central Au+Au collisions at the same \Npart $ $ are also under-predicted by the curves in fig. \ref{Fig2}.}

In fig. \ref{Fig4} we show the ratio of Cu+Cu and Au+Au particle yields where $\langle N_{part} \rangle  =99$. Since the Au+Au yields lack a data point at this value we linearly interpolate between $\langle N_{part} \rangle = 67.5$  and $\langle N_{part} \rangle = 147$. Taking into account the uncertainties, \textcolor{black}{no significant dependence} with respect to strangeness content is observed for the measured data. In addition to the relation in eq. \ref{equ:fraction}, we make comparisons to two other models, EPOS \cite{EPOS1} and AMPT \cite{AMPT}. EPOS is also a core-corona model, however the core-corona splitting is based on the initial energy density, rather than participants that undergo multiple collisions. Other core-corona descriptions have been investigated elsewhere \cite{CCHo}. The AMPT model is based on HIJING \cite{HIJING}, and thus describes particle production in heavy-ion collisions via string excitation and breaking (soft), and mini-jet fragmentation (hard) where the excited nucleons fragment independently. The ratios in the data are better reproduced by EPOS than by AMPT or the parameterisation in eq. \ref{equ:fraction}. However, neither EPOS nor AMPT are able to reproduce individual strange hadron yields in Au+Au and Cu+Cu, as opposed to the ratios of yields between those systems. EPOS is slightly closer to the measured data \cite{HotQuarks}.

In summary, we have presented the enhancement factors for mid-rapidity strange particles as a function of centrality for Cu+Cu and Au+Au \sNN{200} collisions. We have found that the enhancement factors for central Cu+Cu collisions are higher than for mid-central Au+Au collisions with similar numbers of participants. We also found that the qualitative trends for the enhancement factors can be described by a relation that assumes the enhancement factor is proportional to the fraction of participants that undergo multiple collisions. 

We thank Klaus Werner, Joerg Aichelin, Francesco Becattini, and Bin Zhang for discussions, the RHIC Operations Group and RCF at BNL, the NERSC Center at LBNL and the Open Science Grid consortium for providing resources and support. This work was supported in part by the Offices of NP and HEP within the U.S. DOE Office of Science, the U.S. NSF, the Sloan Foundation, the DFG cluster of excellence `Origin and Structure of the Universe'of Germany, CNRS/IN2P3, STFC and EPSRC of the United Kingdom, FAPESP CNPq of Brazil, Ministry of Ed. and Sci. of the Russian Federation, NNSFC, CAS, MoST, and MoE of China, GA and MSMT of the Czech Republic, FOM and NWO of the Netherlands, DAE, DST, and CSIR of India, Polish Ministry of Sci. and Higher Ed., Korea Research Foundation, Ministry of Sci., Ed. and Sports of the Rep. Of Croatia, and RosAtom of Russia.

\bibliography{basename of .bib file}

\end{document}